# Satellites in Discs: Regulating the Accretion Luminosity

D. Syer
*Canadian Institute for Theoretical Astrophysics, MacLennan Labs, 60 St. George Street, Toronto M5S 1A7, Ontario.*

C.J. Clarke
*Institute of Astronomy, Madingley Road, Cambridge CB3 0HA; present address: Queen Mary and Westfield College, Mile End Road, London.*

**ABSTRACT**

We demonstrate, using a simple analytic model, that the presence of a massive satellite can globally modify the structure and emission properties of an accretion disc to which it is tidally coupled. We show, using two levels of numerical approximation, that the analytic model gives reasonable results. The results are applicable to two astrophysical situations. In the case of an active galactic nucleus, we consider the case of a $\sim 10^3 M_\odot$ compact companion to the central black-hole and show that it could modulate the emitted spectrum on a timescale of $\sim 10^5$ years. In the case of a T Tauri accretion disc, a satellite such as a sub-dwarf or giant planet could modify the disc spectral energy distribution over a substantial fraction of the T Tauri star lifetime.

## 1 INTRODUCTION

The theory of tidal interactions between an accretion disc and embedded companion has been elaborated in a series of papers by Lin & Papaloizou (1979a & b, 1986a & b) (see also works by Hourigan & Ward 1984, Goldreich & Tremaine 1980). These papers were primarily concerned with the early solar nebula, and the question of orbital migration of the proto planets, in particular of the giant planets. An annular gap may be formed in the accretion disc due to the action of the tidal forces from the planet, and the formation of such a gap leads to a coupling between the orbit of the planet and the viscous evolution of the disc. In the existing work the satellites were light compared to the local disc, so they migrated on a local viscous timescale. There was little modification of the disc properties outside the gap. In this case, therefore, the presence of the satellite modifies the disc spectral energy distribution only through the removal of a narrow annulus of emitting material around the satellite.

In this paper we explore the effects of a satellite that is massive compared with the local disc. Table 1 gives details of two astrophysical situations in which this could be the case. The inertia of the satellite prevents it from migrating on the local viscous timescale. As a result, disc material is banked up upstream. Downstream of the satellite, material is able to accrete onto the central object, so that the mass of the disc interior to the satellite decreases with time. The satellite therefore causes global changes in the disc structure, and enhances the relative contribution of material upstream to the spectral energy distribution emitted by the disc.

We consider the coupled evolution of a satellite-disc system in the limit that the disc extends only upstream of the satellite. Depletion of the inner disc means that the influence of the inner disc on the satellite dynamics is relatively minor. The neglect of the inner disc implies either that the disc formed from material of higher specific angular momentum than the satellite (as in the case of a collapse scenario where the satellite formed from low angular momentum material, with the disc being formed subsequently from higher angular momentum), or else that the satellite is sufficiently massive for the inner disc to have largely drained onto the central object. In any case, as the satellite migrates radially, it becomes relatively more massive than the local disc, so we anticipate that our calculation of the evolution of the satellite and disc structure is insensitive to the omission of the inner disc. The resultant spectral energy distribution at high frequencies may be more sensitive to this assumption.

The structure of the paper is as follows. In Section 2 we review the process of gap formation and in Section 3 we describe an analytic model for the evolution of the disc and the orbit of the satellite. In Section 4 we describe two numerical approximations to the same problem, and compare their results with the analytic predictions. In the final section we discuss some astrophysical applications of our models, in active galactic nuclei, and in protostellar discs. In the AGN case we show that presence of massive compact objects in the disc would modulate the emission on a timescale of up to $10^5$ years. In the case of a protostellar disc, we consider a satellite which is a massive giant planet or sub-dwarf. The long timescales for the inward transit ($\sim 10^5$ years) implies that such satellites would be expected to cause a substantial fraction of T Tauri stars to show spectra markedly different from those generated by standard steady-state accretion discs.

## 2  THE EVOLUTION OF COUPLED DISC-SATELLITE SYSTEMS

Suppose the satellite has a mass $M_s$ and its orbit is at a radius $r_s$. Consider a 'collision' between the satellite and a parcel of gas with mass $\mu$. The effect of the tidal torque of the planet on the disc may be thought of by analogy with dynamical friction: the fluid at $r > r_s$ moves slower than the satellite, and hence exerts a drag on it, reducing its angular momentum and tending to drive its orbit to smaller radius. The fluid at $r < r_s$ does the opposite. The effect on the disc of the back reaction to the friction is to push the fluid away from the planet; to decrease the specific angular momentum of the fluid at small radii, and to increase it at large radii. The response of the disc is determined by a competition between these tidal perturbations from the satellite, tending to open up an annular gap around the satellite, and pressure and viscous forces in the disc, tending to close the gap.

We will derive an expression for the torque using the analogy with dynamical friction. This is essentially the calculation done by Lin & Papaloizou (1979a), but in the present case, since our purpose is illustrative, we omit constants of order unity. Lin & Papaloizou (1986a) examine the validity of the impulse approximation numerically. (See also Lin & Papaloizou 1986b for a discussion of the same point.) Consider a single collision with a fluid particle at an impact parameter $p$ in the impulse approximation (e.g. Binney & Tremaine 1987). If the relative velocity of the satellite and fluid is $v$, then the angular momentum transferred to the satellite is

$$\Delta L \sim \mu \frac{G^2 M_s^2 r_s}{p^2 v^3} \tag{1}$$

where $G$ is the gravitational constant. Averaging over such impacts, we find that the rate of change of $L$ is

$$\frac{dL}{dt} \sim \int \Delta L \frac{\Sigma}{\mu} v \, dp \sim \int \Sigma \frac{G^2 M_s^2}{p^2 v^2} r_s \, dp, \tag{2}$$

which we write as

$$\frac{dL}{dt} \sim \int \Sigma \Lambda_T r_s \, dp \tag{3}$$

with the torque per unit mass of the disc given by

$$\Lambda_T = \frac{G^2 M_s^2}{p^2 v^2}. \tag{4}$$

Writing $q$ for the mass ratio of the satellite to central object, $M_s/M_*$, and expanding $v$ in a Taylor series about $p = 0$ we have

$$\Lambda_T = q^2 \Omega_s^2 r_s^2 \left(\frac{r_s}{p}\right)^4 \tag{5}$$

where $\Omega_s$ is the angular velocity of the satellite, assuming a Keplerian rotation law and circular orbits for the satellite and disc.

The specific viscous torque near the edge of the gap, assuming that the radial scale length $\sim \Delta$, is

$$\Lambda_\nu \sim \nu \frac{\Omega r}{\Delta} \tag{6}$$

where $\nu$ is the viscosity in the disc and $\Omega$ is the local angular velocity. The radial extension of the gap, $\Delta$, can be evaluated approximately by equating $\Lambda_\nu$ and $\Lambda_T$ with $p \sim \Delta$:

$$\left(\frac{\Delta}{r}\right)^3 \sim \frac{\Omega r^2}{\nu} q^2. \tag{7}$$

Note that in the limit $\nu \to 0$ equation (7) correctly predicts that $\Delta \to \infty$, since in this case there is no equilibrium gap: the satellite continually eats into the disc, at a slower rate as the gap grows. Following Lin & Papaloizou, we define a dimensionless measure of the gap width

$$A = \frac{\Omega r^2}{\nu} q^2 \tag{8}$$

which is the ratio of the viscous to orbital timescales, multiplied by the square of the mass ratio.

Two criteria need to be fulfilled in order for the satellite to succeed in evacuating an annular gap in the disc. Firstly there is the requirement that the gap be not re-filled by radial pressure gradients, which means that the disc vertical scale height, $H$, must be less than the width of the gap

$$\Delta > H. \tag{9}$$

Secondly, there is the requirement that the effect of the satellite is to transfer angular momentum, rather than to accrete from the disc. This means that the gap width should be greater than the Roche radius of the satellite:

$$\Delta > R_L = q^{1/3} r_s. \tag{10}$$

This reduces to

$$q > \frac{\nu}{\Omega r^2}, \tag{11}$$

i.e. $q$ must exceed the local inverse Reynolds number. In the relatively cool and inviscid discs which satisfy these criteria, the satellite sits in an annular gap in the disc of width $\Delta$, exchanging angular momentum with the disc material through the action of tidal torques concentrated near the gap's edge.

In principle it is possible to arrange interior and exterior disc material such that the net torque acting on the satellite is zero, but such a situation will not persist once the disc has undergone significant viscous evolution. The depletion of material interior to the satellite by accretion will result in an imbalance between the torques experienced by the satellite. The net angular momentum transfer is outward from the satellite, which therefore spirals in.

Let us define the term 'reference disc' to mean the steady state disc in the absence of the satellite. We will refer to quantities in the reference disc with the subscript 0. Now we define a dimensionless parameter measuring the mass of the satellite compared to the reference disc

$$B = \frac{4\pi \Sigma_0 r_s^2}{M_s}, \tag{12}$$

where $\Sigma_0$ is the surface density in the reference disc. A relatively light satellite, in the sense that $B \gtrsim 1$, does not substantially modify the surface density of the disc outside the gap. It behaves like a representative fluid element in the disc, and therefore its orbit evolves radially on the viscous timescale (Lin and Papaloizou 1986b).

In this paper we are interested in the opposite case, where $B \lesssim 1$. The disc is unable to push the satellite radially inwards on its own viscous timescale and material therefore forms a reservoir upstream of the gap. Let us define

$$f = \nu \Sigma, \qquad (13)$$

where $\nu$ is the viscosity in the disc. Suppose that the thermal and hydrostatic equilibrium relation between $\nu$ and $\Sigma$ has the power law form given implicitly by

$$\Sigma \propto f^a r^b, \qquad (14)$$

where $a$ and $b$ are constants. The largest contrast in $\Sigma$, over the reference value $\Sigma_0$, allowed by condition (11), is then

$$\frac{\Sigma_\alpha}{\Sigma_0} \sim \left( q \frac{\Omega r^2}{\nu_0} \right)^{a/(1-a)} \qquad (15)$$

where $\nu_0$ is the corresponding reference value of $\nu$. If the surface density behind the satellite rises to $\Sigma_s \sim \Sigma_\alpha$ then the gap will be overwhelmed by viscous diffusion.

In a Shakura-Sunyaev (1973) disc, where the viscosity is given by

$$\nu \sim \alpha \Omega H^2, \qquad (16)$$

with $\alpha$ constant, the criteria (9) and (10) reduce to, respectively,

$$\frac{H}{r} < \left( \frac{q}{\alpha} \right)^{1/2} \qquad (17)$$

and

$$\frac{H}{r} < \left( \frac{q^2}{\alpha} \right)^{1/5}. \qquad (18)$$

If (17) is violated, the gap will be closed by pressure; if (18) is violated, the gap will be closed by accretion onto the satellite. In most cases of astrophysical interest $H/r$ is an increasing function of $r$ in the reference disc, so once the satellite opens up a gap it persists, as long as the disc structure is not modified by the satellite (i.e. if $B > 1$). If the disc structure changes as a result of the presence of the satellite, and $H/r$ becomes a decreasing function of $r$, the values of $q$ and $\alpha$ determine whether it is accretion or pressure which eventually closes the gap. If

$$\left( \frac{q}{\alpha} \right)^{1/2} < \left( \frac{q^2}{\alpha} \right)^{1/5}, \qquad (19)$$

then if the gap closes, accretion will be responsible. This is illustrated in Figure 1 in which we have defined three important radii: $r_o$ is the radius at which the criterion (10) is satisfied; $r_B$ is the radius at which $B = 1$; and $r_c$ is the radius at which (10) is no longer satisfied in the perturbed disc. Figure 1 is entirely schematic, and no particular viscosity law has been used to calculate $H/r$. When $r_s < r_o$ a gap is opened in the reference disc, and when $r_s < r_B$ the disc is banked up behind the satellite.

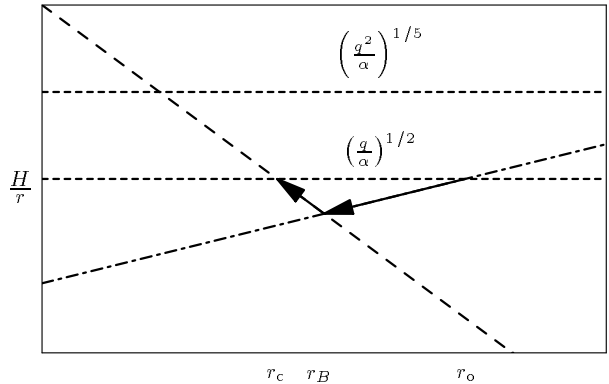

**Figure 1.** The aspect ratio $H/r$ of the disc as a function of $r$. The dot-dashed line is the reference disc. The long-dashed line shows the effect of a satellite in banking up the disc. The lower and upper short-dashed lines are the critical values for closure of the gap by accretion and pressure respectively. The solid line is a possible trajectory of $H/r$ immediately behind the satellite as it migrates radially.

When $r_s < r_c$ or $r_s > r_o$ the satellite cannot maintain an annular gap in the disc, so that the analysis involved in the present work (which assumes tidal coupling between satellite and disc across such a gap) is invalid. Knowledge of the evolution of a satellite embedded in a disc flow requires two-dimensional hydrodynamic calculations beyond the scope of the present work. Thus the following remarks are necessarily speculative. Once the satellite reaches the radius $r_c$, we would expect that the inward flow of material as the gap is overwhelmed would result in a burst of accretion luminosity. The fraction of this material that is accreted by the satellite, rather than by the central object, is however not clear. It is also not clear whether the satellite would remain embedded in the disc thereafter or whether the overflow of a small fraction of the dammed up mass would allow the gap to re-open. In the latter case the satellite might continue to spiral in, maintaining a state of marginal stability against gap closure, with intermittent overflow of the material upstream.

The maximum energy which could be released when the gap is closed is

$$E_b \sim \frac{G M_* \Sigma_c r_c^2}{r_*}. \qquad (20)$$

This corresponds to accretion of the mass $\Delta M \sim \Sigma_c r_c^2$ by the central object of mass $M_*$ and radius $r_*$.

The expectations that a satellite would move as a representative fluid element in the case $B > 1$ was confirmed by Lin and Papaloizou (1986b) using one-dimensional numerical calculations. In their calculation, the usual viscous diffusion equation for the disc was supplemented by an additional tidal torque term (equation 5). The main motivation for these calculations was to evaluate migration timescales for the proto-Jupiter. The question of substantial damming up of the disc by a satellite considerably more massive than the local disc was not extensively investigated. In the following section we consider this question in some detail.

**Table 1.** The approximate values of relevant physical parameters in a selection of astrophysical situations.

|  | $M_*$ | $M_s$ | $r_B$ | $t_B$ |
|---|---|---|---|---|
| AGN | $10^6 M_\odot$ | $10^3 M_\odot$ | $\sim 10^4 GM/c^2$ | $10^5$ y |
| T Tauri star | $1 M_\odot$ | $10^{-2} M_\odot$ | $\sim 100 R_\odot$ | $10^5$ y |

$M_*$ is the central mass; $M_s$ is the satellite mass; $r_B$ is the radius within which the satellite is more massive than the local disc; $t_B$ is the radial migration timescale of the satellite at $r_B$.

## 3 AN ANALYTIC MODEL FOR SATELLITE-DISC EVOLUTION

In this section we first describe the characteristics of the reference disc. We then develop a model for the surface density of the disc when influenced and enhanced by a satellite with $B < 1$ (equation 12). The slow radial drift of the satellite enables viscous torques in the disc to communicate the presence of the satellite to larger radii. We characterise the upstream structure of the disc by dividing it into two sections: an outer region where the influence of the satellite has not yet been felt, and one in the vicinity of the satellite. The boundary between these regions is denoted $r_t$. The outer region, $r_t < r$, is assumed to be in a quasi-steady state with a mass flux determined by the fuelling rate at infinity. The inner region, $r_s < r < r_t$, is assumed to be in a quasi-steady state the properties of which are determined by the angular momentum transfer between the satellite and the inner edge of the disc. This results in a lower mass flux in the inner region, on account of the damming effect of the satellite. The limit of $r_t \to \infty$ is analytic and explicit.

### 3.1 The reference disc

In a thin, Keplerian accretion disc with no external torques, $f$ is determined by

$$\frac{\partial \Sigma}{\partial t} = \frac{3}{r} \frac{\partial}{\partial r} r^{\frac{1}{2}} \frac{\partial r^{\frac{1}{2}} f}{\partial r} \qquad (21)$$

(e.g. Pringle 1981), and so in a steady state

$$\frac{\partial}{\partial r} r^{\frac{1}{2}} \frac{\partial r^{\frac{1}{2}} f}{\partial r} = 0. \qquad (22)$$

In a steady-state disc we assume an outer boundary condition

$$2r^{\frac{1}{2}} \frac{\partial r^{\frac{1}{2}} f}{\partial r} = \frac{\dot{M}_0}{3\pi} = f_\infty, \quad r \to \infty, \qquad (23)$$

where $\dot{M}_0$ is the mass flux through the disc. At the inner boundary, $r_*$, we assume the torque on the inner edge of the disc vanishes, whence

$$f = 0, \quad r = r_*. \qquad (24)$$

The general solution to (22) is

$$f = X + \frac{Y}{r^{\frac{1}{2}}} \qquad (25)$$

where $X$ and $Y$ are constants. With the boundary conditions (23) and (24) we find that

$$f = f_0 = f_\infty - f_\infty \left(\frac{r_*}{r}\right)^{\frac{1}{2}}. \qquad (26)$$

In the limit $r_* \to 0$, which is adequate for our purposes, $f_0 = f_\infty$.

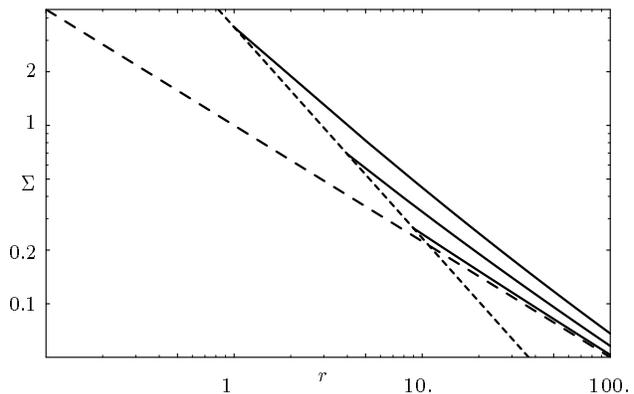

**Figure 2.** The surface density $\Sigma_s$ behind the satellite as a function of $r_s$ (short dashed line), in the analytic model with $r_t \to \infty$. The constants of the model are chosen so that $B = 0.04$ at $r = 1$. The solid lines are snapshots of the surface density $\Sigma$. Also shown is the reference disc (long dashed line).

### 3.2 The satellite

We consider the situation in which the disc interior to $r_s$ is empty ($\Sigma = 0$). Such a situation would arise if the disc was formed in this way, but would also be approximately realised in the limit of a sufficiently massive satellite: the satellite acts as a dam, allowing the inner disc to drain away in a time much shorter than the migration timescale of the planet.

We assume that the tidal coupling between the satellite and the disc is confined to a very narrow region at the inner edge of the gap. In this case equation (21) adequately describes the evolution of $f$. The outer boundary condition remains the same, on the assumption that far enough upstream of the satellite the disc must approach the undisturbed solution. The inner boundary condition is determined by supposing that a) the satellite and disc edge move with a common velocity, and b) there is overall conservation of angular momentum. The radial velocity in the disc is given by conservation of mass as

$$v = -\frac{3}{\Sigma r^{\frac{1}{2}}} \frac{\partial r^{\frac{1}{2}} f}{\partial r} \qquad (27)$$

Evaluating this at $r = r_s$, and equating it to the velocity of the satellite, $v_s$, provides the inner boundary condition. (Provided $\Delta$ remains considerably smaller than $r_s$, then $v$ and $v_s$ cannot substantially differ.)

Angular momentum conservation is guaranteed by equation (21) everywhere except at the inner edge of the gap. The viscous torque on the edge of the disc is

$$G = 3\pi f_s r_s^2 \Omega_s \tag{28}$$

assuming a Keplerian rotation law. The rate of change of the angular momentum of the satellite is

$$\frac{d}{dt} M_s r_s^2 \Omega_s = \frac{1}{2} M_s r_s^2 \Omega_s \frac{v_s}{r_s} \tag{29}$$

again assuming a Keplerian rotation law. Conservation of angular momentum is thus expressed as

$$3\pi f_s r_s^2 \Omega_s + \frac{1}{2} M_s r_s^2 \Omega_s \frac{v_s}{r_s} = 0, \tag{30}$$

so the value of $v$ in the boundary condition (27) is

$$v_s = -\frac{6\pi r_s f_s}{M_s}. \tag{31}$$

We write equation (31) as

$$v_s = -B \frac{f_s}{f_0} \frac{r_s}{t_0}, \tag{32}$$

where

$$t_0 = \frac{2}{3} \frac{r_s^2}{\nu_0} \tag{33}$$

is the viscous timescale in the undisturbed disc.

In reality the disc will never reach a global steady state, since $v_s$ is always non-zero. But, for small $B$, the satellite moves slowly compared to the local viscous timescale, so we make one more approximation: that the disc is able to reach a steady state for all values of $r_s$. In the outer region of the disc, $r_t < r$, we suppose that it has not had time to adjust to the presence of the satellite, and thus that

$$f = f_0 = f_\infty, \qquad r_t < r. \tag{34}$$

In the inner region, $r_s < r < r_t$, we solve (22) with the boundary conditions (34) at $r = r_t$, and a combination of (27) and (32) at $r = r_s$. Using (14) and (25), we find that the required solution to (22) is

$$f = f_a + (f_s - f_a)\left(\frac{r_s}{r}\right)^{\frac{1}{2}} \tag{35}$$

where $3\pi f_a < \dot{M}_0$ is the mass flux through the inner disc. Equating (34) and (35) at $r = r_t$ we find that

$$f_a + (f_s - f_a)\left(\frac{r_s}{r_t}\right)^{\frac{1}{2}} = f_\infty, \tag{36}$$

with

$$f_s = f_0 \left(\frac{f_0}{f_a} B\right)^{-1/(1+a)}. \tag{37}$$

Once $r_t$ is specified, equations (36) and (37) are sufficient to determine $f_s$ and $f_a$, given $r_s$.

The model assumes that the disc is in a steady state everywhere apart from at $r_t$. At $r_t$ the mass fluxes in the inner and outer disc are unbalanced, so mass is not conserved unless $r_t$ moves at just the right rate. The evolution of $r_t$ would be specified by equations (36) and (37), together with an integro-differential equation that ensured conservation of mass. We do not develop the latter in this paper, but retain the notion of an inner and outer disc for the purpose of qualitatively understanding our numerical simulations.

### 3.3 The limit $r_t \to \infty$

In the limit $r_t \to \infty$, the model becomes explicit and analytic. We expect that this limit should be approached as $B \to 0$, in which the satellite moves very slowly, and is able to influence the surface density very far upstream. In this limit we find that

$$f_s = f_0 B^{-1/(1+a)}. \tag{38}$$

This is an approximate description of the 'banking up' of the surface density behind the satellite. The surface density just behind the satellite is given by

$$\Sigma_s = \Sigma_0 B^{-a/(1+a)}. \tag{39}$$

Provided that $a > 0$ we are reassured that the smaller the value of $B$, the greater is the surface density $\Sigma_s$. The surface density can thus be enhanced so that $\Sigma_s \geq \Sigma_0$. This is illustrated in Figure 2, in which we have used a viscosity law (equation 14) with $a = 0.65$ and $b = -0.65$, which is appropriate to quiescent protostellar discs (Clarke, Lin & Pringle 1990). In all the quantitative work in this paper, and in all the figures (except Figure 1), we use the same viscosity law.

Combining (38) and (32) we can find $v_s$ explicitly in terms of $r_s$:

$$v_s = -B^{a/(1+a)} \frac{r_s}{t_0}. \tag{40}$$

Integrating this equation we find the evolution of $r_s$ with time

$$r_s = r_i \left[1 - c \frac{t}{t_i} B_i^{a/(1+a)}\right]^{1/c}, \tag{41}$$

where $t_i$ and $B_i$ are the values of $t_0$ and $B$ evaluated at $r = r_i$, and

$$c = \frac{2+b}{1+a}. \tag{42}$$

Comparing equation (40) to the radial velocity at $r_s$ in the undisturbed disc

$$v_0 = -\frac{r_s}{t_0}, \tag{43}$$

we find that

$$v_s = v_0 B^{a/(1+a)}. \tag{44}$$

For $B < 1$, $a > 0$, $b > -2$, we see that $v_s < v_0$. Our assumption of a quasi-steady state must break down when $v_s \gtrsim v_0$—but we expect $v_s$ to be at most $v_0$, when $B \sim 1$.

## 4 NUMERICAL MODELS

### 4.1 Infinitesimal boundary layer

To test the steady-state assumption in the analytic model, we now drop that assumption, but retain the inner boundary condition (equations 27 and 31), and then solve equation (21) numerically. The righthand side of equation (21) is differenced in a standard implicit scheme, and a Lax average is employed for the lefthand side (e.g. Press *et al.* 1992). An iterative procedure is employed for solving the implicit equations, as described for example by Potter (1987). A uniform grid in the variable $r^{1/2}$ is employed, with 6400 points in the range $r = r_{\rm in} = 0.25$ to $r = r_{\rm out} = 1024$. The initial radius of the satellite orbit is $r = 6$, and the inner boundary condition is applied at the nearest gridpoint upstream. At every timestep, the new location of the satellite is computed from equation (31), and if it crosses a gridpoint, the inner boundary condition is applied at that gridpoint.

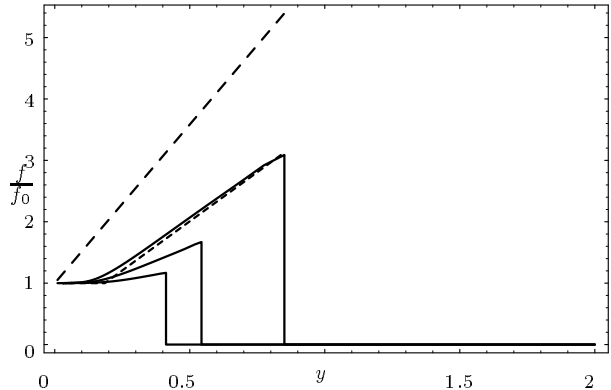

**Figure 3.** The result of dropping the steady-state assumption: snapshots of $f(y)$ with $B = 0.04$ at $r = 1$ (solid curves). The short-dashed curve shows the analytic prediction for the right-most snapshot with $r_t = 30$; the long-dashed curve shows the analytic prediction with $r_t \to \infty$.

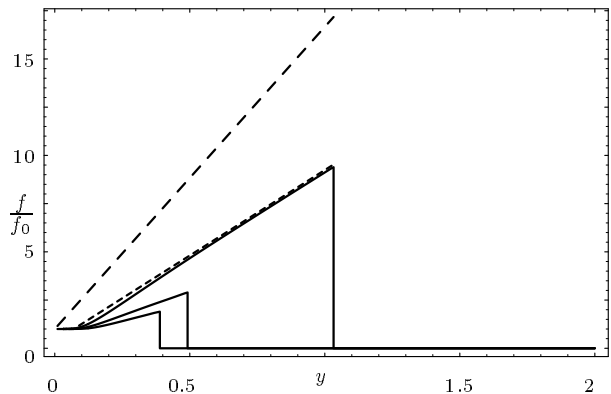

**Figure 4.** As Figure 3 but with $B = 0.01$ at $r = 1$. The short-dashed curve in this case is with $r_t = 200$ for the right-most snapshot; the long-dashed curve shows the analytic prediction with $r_t \to \infty$.

In Figure 3 we show two snapshots of the disc structure for the case $B = 0.04$ at $r = 1$; we plot $f/f_0$ as a function of $y = r^{-1/2}$. The figure illustrates that the disc is indeed well represented by a sequence of quasi-equilibrium states—$f(y)$ can be approximated by two straight line sections. The radius of the intersection of these straight lines, $r_t$, does not change much as the satellite migrates radially. (We have checked the independence of $r_t$ on the position of the outer boundary by carrying out additional calculations with $r_{\text{out}} = 256$, in which the results were the same.) Figure 4 ($B = 0.01$ at $r = 1$) shows the case of a heavier satellite. A more massive satellite causes a larger surface density enhancement and is able to influence the disc out to a larger radius. The analytic model is in each case able to fit the numerical results adequately, once $r_t$ is specified.

### 4.2 Explicit torque

As a second numerical comparison with the analytic model we drop the assumption of an infinitesimal boundary layer, and include the tidal torque from the satellite explicitly. Thus equation (5) appears as a source term in the diffusion equation (21):

$$\frac{\partial f}{\partial t} = \frac{1}{r}\frac{\partial}{\partial r}\left(3r^{\frac{1}{2}}\frac{\partial r^{\frac{1}{2}}f}{\partial r} - \frac{\Lambda_T \Sigma}{\Omega}\right) \tag{45}$$

(Lin & Papaloizou 1986b). For practical purposes, $\Lambda_T$ has to be softened on a scale which is smaller than $\Delta$. We use the prescription of Lin & Papaloizou (1986b), but since $\Sigma$ vanishes in the gap anyway, the details of the softening are not important once the gap has formed.

Angular momentum conservation is ensured by setting the rate of change of the angular momentum of the planet equal to the integrated contributions of the tidal torque exerted by the disc:

$$\frac{\mathrm{d}}{\mathrm{d}t}\left(M_s \Omega_s r_s^2\right) = -\int \Lambda_t \Sigma \, 2\pi r\,\mathrm{d}r. \tag{46}$$

The numerical resolution has to be high for small $A$ (small gap width). Here we used 12800 grid elements in the radial range $r \in [0.25, 256]$. The results of the explicit torque calculation are shown in Figure 5. In common with Figure 3, we see that $f$ is approximately described by two straight line segments as in the analytic model, and that the location of $r_t$ does not change much with time. This illustrates that, as expected, the form of the torque distribution has negligible effect on the disc structure at large distances upstream. The disc structure close to the satellite is somewhat different when the distributed torque is included. The disc is effectively obeying the boundary condition (equation 31) at a radius where the tidal torque is maximum. But the tidal torque does not immediately drop to zero outside this radius, and the surface density can continue to rise. As a result, the peak surface density is greater than that implied by equation (31). The distributed torque model predicts higher peak densities than the infinitesimal boundary layer model.

In Figure 6 we compare the results of using two different values of $A$ (equation 8). A smaller value (Figure 6a) leads to a narrower gap, and a sharper turn over of the surface density profile at the gap edge, but does not significantly affect the value of the surface density at the turn over. We conclude that the gap width ($\Delta \sim A^{1/3} r_s$) has a minor effect on the surface density enhancement, which is controlled largely by the ratio of the disc mass to the mass of the satellite ($B$).

### 4.3 Summary

Figure 7 shows a comparison of the evolution of $r_s$ for the numerical models with the analytic model. In the case of the infinitessimal boundary layer model, the surface density enhancement is smaller than predicted by $r_t \to \infty$. The surface density enhancement determines the radial evolution of the satellite through equation (31), so the evolution of $r_s$ is slower than the analytic prediction. In the case of the explicit torque model it is slower because the radius at which equation (31) is satisfied is closer to the satellite than the peak surface density.

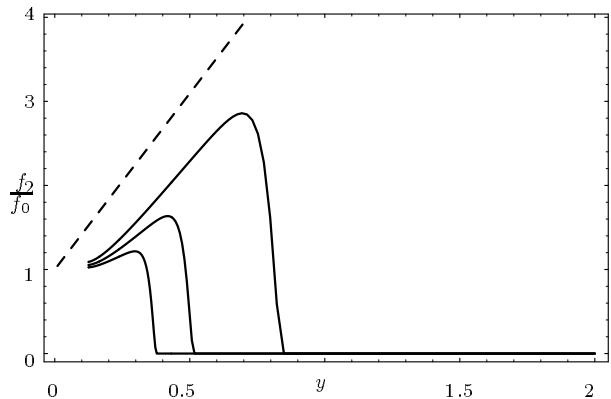

**Figure 5.** The result of including an explicit tidal torque: the explicit torque model with $B = 0.04$ at $r = 1$ and $A = 0.25$. The dashed curve shows the analytic prediction for the right-most snapshot with $r_t \to \infty$.

The numerical results show that the disc has a quasi-equilibrium structure with $r_t$ varying slowly. The limit $r_t \to \infty$, in which all the equations are analytic and explicit should be approached in the limit of a massive satellite. Since the numerical calculations are considerably more expensive in effort and computational resources, the advantages of the analytic approach are obvious.

## 5 DISCUSSION

We have derived analytic expressions for the surface density enhancement in a disc which is modified by the presence of a massive satellite at its inner edge (equations 12 and 39). We have verified the accuracy of these expressions, to order unity, using two levels of numerical approximation.

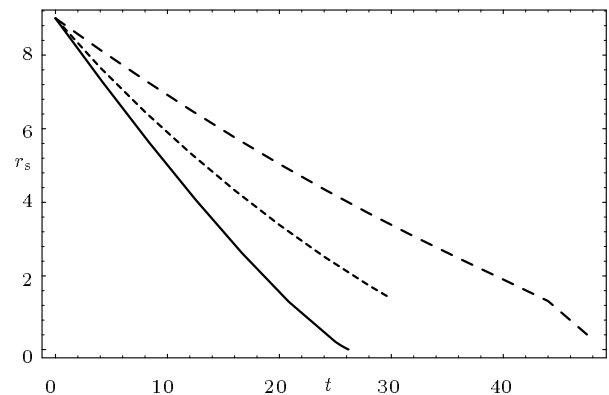

**Figure 7.** The evolution of $r_s$ as a function of time (in units of the viscous time $t_0$ at $r = 1$), for $B = 0.04$ at $r = 1$. The long-dashed line is the explicit torque model; the short-dashed line is the infinitessimal boundary layer model; and the solid line is the analytic model with $r_t \to \infty$.

A satellite initially at large radius, where $B > 1$, moves likes a representative fluid element and scarcely modifies the structure of the disc. Provided $B$ decreases with decreasing radius ($b > -2$, equation 14), such a satellite may eventually reach the radius, $r_B$, at which $B = 1$. The disc surface density is subsequently enhanced behind the satellite because it is relatively massive compared with the local disc ($B < 1$).

In Figure 8 we compare the spectral energy distribution, $F_\nu$, in a disc dammed up behind a satellite with that generated by a reference disc with a zero torque boundary condition at its inner edge. In all cases we assume that the disc radiates everywhere as a black body with local effective temperature fixed by the local viscous dissipation rate. When the satellite is at radii considerably larger than the inner edge of the reference disc, the assumed absence of ma-terial interior to the satellite implies that the spectral energy distribution is deficient in higher frequency radiation compared with the reference disc. (This effect would be more marked if the contribution from boundary layer emission were included—in the case that the reference disc accreted onto a stellar surface at its inner edge). As the satellite progresses to smaller radii, the emission is enhanced over a wide range of frequencies: the energy lost by the inward spiralling satellite is radiated through the disc, and the enhanced emission from just upstream of the satellite compensates for the absence of material interior to it. The spectrum of a classic steady state accretion disc may be characterised as a power law $F_\nu \propto \nu^{1/3}$ over a wide dynamic range. When a satellite with $B < 1$ is present this changes to $F_\nu \propto \nu^{5/7}$ (see Appendix for derivation).

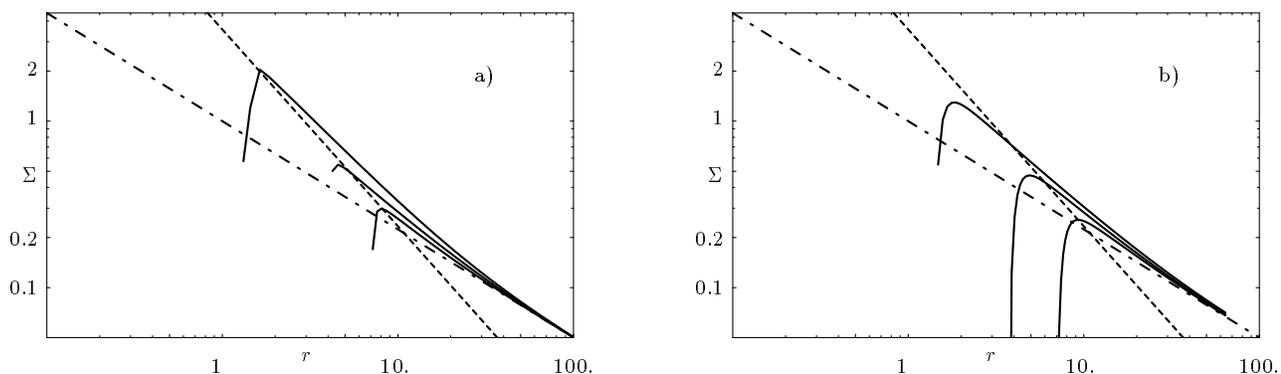

**Figure 6.** Snapshots of the surface density in the explicit torque model with $B = 0.04$ at $r = 1$ and a) $A = 0.0025$, b) $A = 0.25$. The dashed curves show the analytic prediction for $\Sigma_s$ with $r_t \to \infty$.

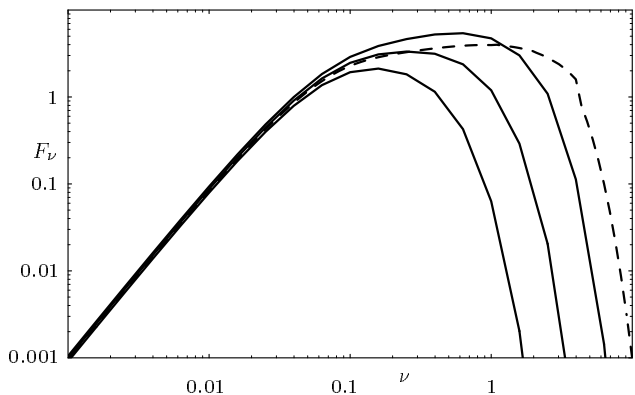

**Figure 8.** The evolution of the spectral energy density $F_\nu$ from the accretion disc, as a satellite drifts radially inwards. The dashed line shows $F_\nu$ for the reference disc, with inner radius $r_* = 0.5$, $r_{\rm out} = 200$ and $a = -b = 3/5$. The solid lines show $F_\nu$ from the banked-up disc, which is assumed to be empty inside the radius of the orbit of the satellite. From left to right, the solid curves are for $r_s = 20, 10$ and $5$. In Each case $r_t \to \infty$ and $B = 0.01$ at $r = 1$.

We consider two applications of astrophysical interest: a disc around a pre-main sequence (T Tauri) star with a sub-dwarf satellite; and a disc in an Active Galactic Nucleus where the satellite is either of stellar mass or is a compact object of mass $\sim 10^3 M_\odot$ (see Syer, Clarke and Rees 1991 and Lin 1994 for arguments relating to the possible presence of such objects in AGN discs). The T Tauri model uses a central mass of $1 M_\odot$, viscous alpha parameter $10^{-3}$ and an accretion rate of $10^{-7} M_\odot {\rm y}^{-1}$ (details of the disc structure in Clarke, Lin and Pringle 1990). The AGN model uses a central mass of $10^6 M_\odot$, viscous alpha parameter $10^{-1}$ and an accretion rate of one tenth the Eddington rate (details of the disc structure in Clarke 1987, 1988).

For a stellar mass object in an AGN disc, the disc overwhelms the gap created by the satellite at effectively all radii. When the mass of the satellite is $10^3 M_\odot$, and in the T Tauri model the gap opening criteria are satisfied and it is meaningful to calculate $r_B$, as listed in Table 1. A massive compact satellite in an AGN disc would bank up the disc behind it for $r_s < r_B \sim 10^4$ times the radius of the central object (at this radius the disc is cool and unionised). In the pre-main sequence case, a sub-dwarf of mass $10^{-2} M_\odot$ would bank up the surface density in the disc for $r_s < r_B \sim 100 R_\odot$. The time taken for the companion to spiral in from $r_B$ is of order the disc's viscous timescale at $r_B$; $t_s$, the time to spiral in from a smaller radius is greater than the local reference viscous timescale, $t_0$, because the inertia of the satellite is large compared with the local disc. At smaller radii the ratio $t_s/t_0$ increases. However, $t_B$ provides a measure of the interval of time over which the satellite is able to significantly modify the structure and spectral energy distribution of the disc. Thus a massive black hole in an AGN disc would give rise to deviations from classic steady-state disc emission during its transit of the inner disc. It may also affect the continuum emission, on the assumption that the continuum arises from physical processes near the central mass, and therefore relies on the disc to transport fuel to small radii. A realistic calculation of the spectral energy distribution would also have to include the remnant disc interior to the satellite, and would require knowledge of the physical processes that give rise to the continuum emission. We find that for a large range of AGN disc parameters, the disc is unable to close the annular gap created by the companion and that the banking up of the disc behind the satellite (and associated increase in luminosity) continues until the companion reaches the last stable orbit and is swallowed by the central black hole. At this point the spectral energy distribution of the disc reverts to that of the usual steady state disc on a timescale comparable with $t_B$.

In the pre-main sequence case the corresponding value of $t_B$ is long ($\sim 10^5$ years) so that the disc structure and spectrum would deviate from its steady state values over a considerable period. Although, as above, the calculation of the spectral energy distribution should also contain a component from the disc inside the satellite (a contribution that becomes increasingly unimportant once the satellite is well inside $r_B$), Figure 8 gives some indication of the types of spectral energy distributions that might be generated. Thus, contrary to previous attempts to model the effect of satellites on disc emission merely in terms of "missing emission" from annular gaps (e.g. Marsh and Mahoney 1992), we find that satellites massive compared with the local disc produce global modifications in the spectral energy distribution: the disc emission is enhanced upstream of the satellite and this produces a variety of signatures in the resultant spectral energy distribution. While the satellite is at relatively large radius, the increased strength of the outer disc produces a hump in emission at low frequencies. This is in line with the observed tendency of T Tauri stars to exhibit too much power at low frequencies compared with standard steady-state disc models. Once the satellite is very close to the star, its effect is to *steepen* the spectral energy distribution from the classic $\nu^{1/3}$ to a $\nu^{5/7}$ power law. This effect would be masked observationally by the contribution from the boundary layer between the depleted disc and the central object.

In the case of a satellite in a T Tauri disc we find that the disc is never able to overwhelm the gap created by the satellite, provided it remains cool and largely unionised; if this is the case, the satellite continues to spiral inwards until it is either enveloped or tidally disrupted by the central star. Another possibility is that the temperature in the disc upstream of the satellite reaches the threshold for the hydrogen ionisation; if this is the case then the associated rapid rise in the viscosity could enable the disc to overwhelm the gap and flow rapidly inwards past the satellite. Such a scenario may offer a mechanism for the generation of FU Orionis outbursts in pre-main sequence stars (Clarke & Syer in preparation).

## ACKNOWLEDGMENTS

CJC is grateful for the hospitality of the MPA, Garching where this work was initiated. We are indebted to Drs F. and E. Meyer for input during the early stages of this project, and, in particular, for suggesting the approach adopted in Section 4.1. DS acknowledges financial support from PPARC and NSERC of Canada.

## REFERENCES

Binney, J., and Tremaine, S., 1987, *Galactic Dynamics*,
  Princeton: Princeton University Press.
Clarke, C.J. 1987, *Thesis*, Oxford.
Clarke, C.J. 1988, MNRAS, 235, 881
Clarke, C.J., Lin, D.N.C., and Pringle, J.E. 1990, MNRAS, 242,
  439
Goldreich, P., and Tremaine, S. 1980, ApJ, 241, 425
Herbig, G.H. 1989, *ESO Workshop*, 33, 233 (ed
B. Reipurth)
Hourigan, K., and Ward, R.W. 1984, Icarus, 60, 29
Lin, D.N.C., 1994, in *Theory of Accretion Disks 2*, eds
  W.J.Duschl, Dordecht:Kluwer.
Lin, D.N.C., and Papaloizou, J. 1979a, MNRAS, 186, 799
Lin, D.N.C., and Papaloizou, J. 1979b, MNRAS, 188, 191
Lin, D.N.C., and Papaloizou, J. 1986a, ApJ, 307, 395
Lin, D.N.C., and Papaloizou, J. 1986b, ApJ, 309, 846
Marsh, K.A., and Mahoney, M.J. 1992, ApJ, 395, L115
Potter, M.C. 1987, *Mathemetica Methods in the Physical
  Sciences*, Englewood Cliffs, N.J.: Prentice-Hall.
Press, W.H., Teukolsky, S.A., Vetterling, W.T., and Flannery,
  B.P. 1992, *Numerical Recipes*, Cambridge: Cambridge
  University Press.
Pringle, J.E. 1981, ARAA, 19, 137
Shakura, N.I., and Sunyaev, R.A. 1973, AA, 24, 337

## APPENDIX : THE EMITTED SPECTRUM OF THE DISC

The response of the disc to the tidal forces from the satellite causes addtional viscous dissipation in the disc. If the inner disc is evacuated, then there is also a deficit of high frequency radiation compared with the reference disc. Here we calculate the spectral energy density of the radiation from the disc in our analytic model. We show that the former effect dominates at late times, and the latter dominates when the satellite first begins to stem the radial flow in the disc.

On the assumption that the disc is optically thick, at each radius there is an effective temperature of the emitted radiatio $T$ given by

$$\sigma T^4 \propto \frac{f}{r^3} \qquad (47)$$

where $\sigma$ is the Stefan-Boltzmann constant (e.g Pringle 1981). The spectrum from each annular element of the disc is the blackbody spectrum

$$B_\nu(T) \propto \frac{\nu^3}{\exp(h\nu/kT) - 1} \qquad (48)$$

where $k$ is Boltzmann's constant. Thus the spectrum of the disc as a whole is

$$F_\nu \propto \int_{r_{\rm in}}^{r_{\rm out}} B_\nu(T)\, 2\pi r\, dr \qquad (49)$$

where $r_{\rm in}$ and $r_{\rm out}$ are the inner and outer radii of the disc, respectively.

Suppose that

$$T = T_* \left(\frac{r_{\rm in}}{r}\right)^\beta \qquad (50)$$

in which case for $h\nu \ll kT_*$ we can rewrite equation (49) as

$$F_\nu \propto T_*^{2/\beta} \nu^{3-2/\beta} \int_0^{x_{\rm out}} \frac{x^{1+2/\beta}\, dx}{e^x - 1} \qquad (51),$$

where $x = h\nu/kT$ and $x_{\rm out} = h\nu/kT_{\rm out}$. For frequencies such that $kT_{\rm out} \ll h\nu \ll kT_*$ we may take $x_{\rm out} \gg 1$ and hence $F_\nu \propto \nu^{3-2/\beta}$.

In the reference disc $\beta = 3/4$ and we recover the classic result that $F_\nu \propto \nu^{1/3}$. When the disc is truncated by a satellite we take $r_{\rm in} = r_s$, and assume that the disc interior to the satellite is empty. For $B \ll 1$, equations (35) and (38) imply that

$$f \sim B^{-1/(1+a)} \left(\frac{r_s}{r}\right)^{1/2}. \qquad (52)$$

Thus we find that $\beta = 7/8$ and

$$F_\nu \propto B^{-4/7(1+a)} \nu^{5/7} \qquad (53)$$

Thus, the spectrum of the disc is steeper than the reference disc, and the normalisation increases as $B$ decreases (i.e. as $r_s$ decreases) for $a > -1$.

This paper has been produced using the Blackwell Scientific Publications TEX macros.